\documentclass[pdflatex,sn-mathphys]{sn-jnl}%
\usepackage{hyperref}
\usepackage{url}
\usepackage{booktabs}
\usepackage{graphicx}
\usepackage{caption}
\usepackage{subfigure}
\usepackage{algorithm}
\usepackage{multirow}
\usepackage{xcolor}
\usepackage{enumitem}
\usepackage{amsmath,amsfonts,bm}

\def\gG{{\mathcal{G}}}
\def\gL{{\mathcal{L}}}

\def\ve{{\bm{e}}}
\def\vh{{\bm{h}}}
\def\vm{{\bm{m}}}
\def\vx{{\bm{x}}}
\def\vy{{\bm{y}}}

\def\mA{{\bm{A}}}
\def\mB{{\bm{B}}}
\def\mE{{\bm{E}}}
\def\mH{{\bm{H}}}
\def\mX{{\bm{X}}}

\newcommand{\eg}{\textit{e}.\textit{g}., }
\newcommand{\ie}{\textit{i}.\textit{e}., }

\jyear{2023}%

\theoremstyle{thmstyleone}%
\theoremstyle{thmstyletwo}%

\theoremstyle{thmstylethree}%

\begin{document}

\title[Protein-Specific Variant Effect Prediction with Graph Neural Networks]{Accurate and Definite Mutational Effect Prediction with Lightweight Equivariant Graph Neural Networks}

\author[1,2]{\fnm{Bingxin} \sur{Zhou}}\email{bingxin.zhou@sjtu.edu.cn}

\author[1]{\fnm{Outongyi} \sur{Lv}}\email{harry\_lv@sjtu.edu.cn}

\author[3]{\fnm{Kai} \sur{Yi}}\email{kai.yi@unsw.edu.au}

\author[1]{\fnm{Xinye} \sur{Xiong}}\email{cintia\_bear@sjtu.edu.cn}

\author[1]{\fnm{Pan} \sur{Tan}}\email{tpan1039@alumni.sjtu.edu.cn}

\author[1,4]{\fnm{Liang} \sur{Hong}}\email{hongl3liang@sjtu.edu.cn}

\author*[1,2,3]{\fnm{Yu Guang} \sur{Wang}}\email{yuguang.wang@sjtu.edu.cn}

\affil[1]{\orgdiv{Institute of Natural Sciences}, \orgname{Shanghai Jiao Tong University}, \orgaddress{ \city{Shanghai}, \postcode{200240}, \country{China}}}

\affil[2]{\orgdiv{Shanghai National Center for Applied Mathematics (SJTU Center)}, \orgaddress{ \city{Shanghai}, \postcode{200240}, \country{China}}}

\affil[3]{\orgdiv{School of Mathematics and Statistics}, \orgname{University of New South Wales}, \orgaddress{\street{Street}, \city{Sydney}, \postcode{2052}, \state{NSW}, \country{Australia}}}

\affil[4]{\orgdiv{School of Physics and Astronomy \& School of Pharmacy}, \orgname{Shanghai Jiao Tong University}, \orgaddress{ \city{Shanghai}, \postcode{200240}, \country{China}}}


\abstract{Directed evolution as a widely-used engineering strategy faces obstacles in finding desired mutants from the massive size of candidate modifications. While deep learning methods learn protein contexts to establish feasible searching space, many existing models are computationally demanding and fail to predict how specific mutational tests will affect a protein's sequence or function. This research introduces a lightweight graph representation learning scheme that efficiently analyzes the microenvironment of wild-type proteins and recommends practical higher-order mutations exclusive to the user-specified protein and function of interest. Our method enables continuous improvement of the inference model by limited computational resources and a few hundred mutational training samples, resulting in accurate prediction of variant effects that exhibit near-perfect correlation with the ground truth across deep mutational scanning assays of 19 proteins. With its affordability and applicability to both computer scientists and biochemical laboratories, our solution offers a wide range of benefits that make it an ideal choice for the community.}


\maketitle

\section{Introduction}
Mutation is a fundamental biological process that involves changes in the amino acid (AA) types of specific proteins. However, the functions of wild-type proteins may not always satisfy bio-engineering needs. Therefore, it is necessary to optimize their function, \ie fitness, through favorable mutations. This operation is essential when designing antibodies \citep{wu2019machine,pinheiro2021metabolic,shan2022deep} or enzymes \citep{sato2019protein,wittmann2021advances}.
A protein typically consists of hundreds to thousands of AAs, where each belongs to one of the twenty AA types. To optimize a protein's functional fitness, a greedy search is conventionally carried out in the local sequence. The process involves mutating AA sites to improve the functionality of the protein, rendering a mutant with a higher gain-of-function \citep{rocklin2017global}. Such a process is called \emph{directed evolution} \cite{arnold1998design}. 

\begin{figure}[!t]
    \centering
    \includegraphics[width=\textwidth]{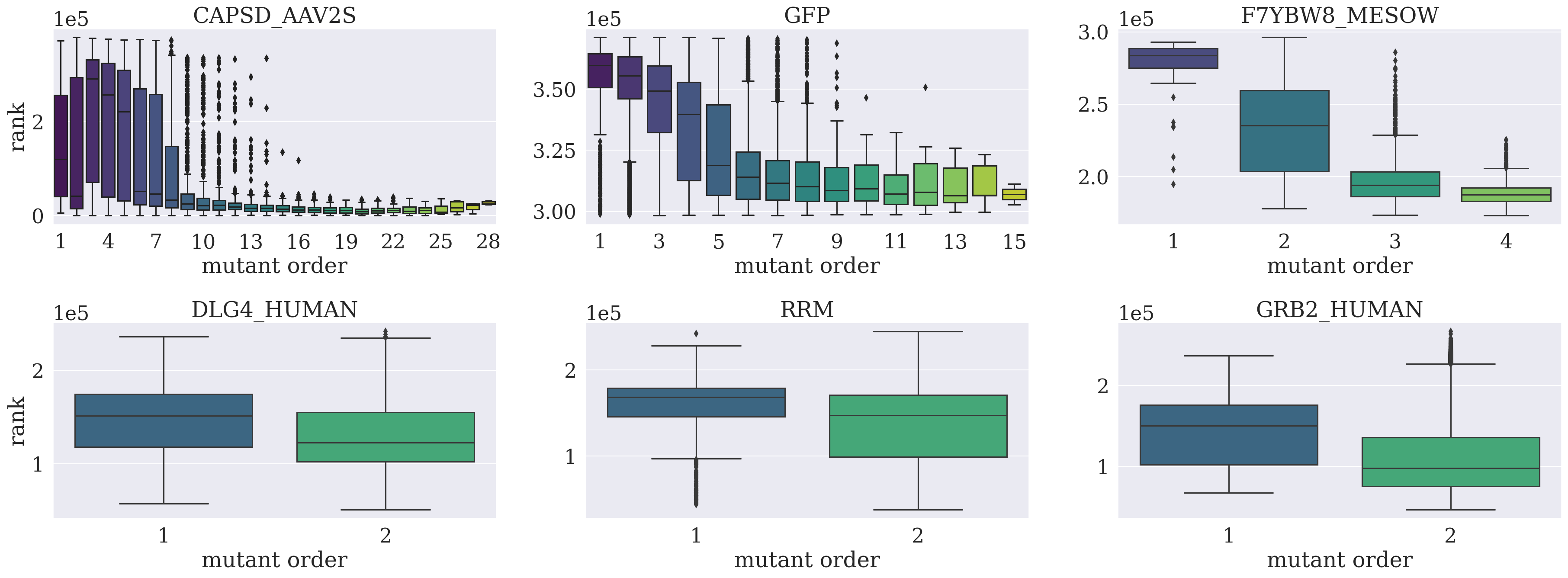}
    \caption{Mutating on more AA sites frequently results in better fitness, \ie smaller average rank values.}
    \label{fig:box:rank}
\end{figure}

Obtaining mutants with high fitness requires mutating multiple AA sites of the protein, known as \emph{deep mutations} (see Fig.~\ref{fig:box:rank}). However, this process incurs significant experimental costs due to the astronomical number of potential mutation combinations. Thus, there is a need for \textit{in silico} examination of protein variant fitness. A handful of deep learning methods have been developed to accelerate the discovery of advantageous mutants. For instance, Lu \textit{et al.} \cite{lu2022machine} applied \textsc{3DCNN} to identify a new polymerase with a beneficial single-site mutation that enhanced the speed of degrading Polyethylene terephthalate (PET) by 7-8 times at 50$^{\circ}\text{C}$. Luo \textit{et al.} \cite{luo2021ecnet} proposed \textsc{ECNet} to predict functional fitness for protein engineering with evolutionary context and guide the engineering of TEM-1 $\beta$-lactamase to identify variants with improved ampicillin resistance. Thean \textit{et al.} \cite{thean2022machine} enhanced SVD with deep learning in identifying Cas9 nuclease variants that possess higher editing activities than the derived base editors in human cells.

The scarcity of labeled protein data and the uniqueness of distinct protein families make it challenging to train supervised learning models directly from observed mutants. As an alternative, researchers frequently pre-train models to encode protein sequences or structures and use the learned protein representations subsequently for specific tasks, such as \textit{de novo} protein design from scratch \citep{hsu2022esmif1}, mutational effect prediction \citep{ingraham2019generative,jing2020learning,meier2021esm1v,notin2022tranception}, and higher-level structure prediction \citep{ahmed2021protbert}. This paper establishes a lightweight supervised learning strategy that trains on a few labeled protein data to suggest favorable mutational directions. Compared to existing methods that incrementally raise the prediction performance, our method is conceivably more suitable for guiding real scientific discovery. For instance, our method substantially surpasses one of the state-of-the-art models \textsc{ESM-if1}~\cite{hsu2022esmif1}on \textbf{RRM} for variant effects prediction. The ground-truth fitness score correlation of our method is higher than $0.9$, whereas the best obtained by \textsc{ESM-if1} barely exceeds $0.4$. 

Similar to training a novice in a new discipline, we maximize the learning efficiency of our proposed model by initially feeding it with a scalable dataset of wild-type proteins for context learning ahead of specifying proteins and functionalities. The pre-training process does not involve any supervision with real learning tasks or the target labels, and it is usually referred to as a \textit{self-supervised learning} procedure. In literature, the problem is commonly reformatted to mini-\textit{de novo} design that infers a specific AA type from its microenvironments, such as its neighboring AA types and local structure. Inferring optimal mutational directions from wild-type proteins can be viewed as a simulation of natural selection, given that mutation in natural conditions involves random changes of the AA type toward any of the other $19$ AA types. It is suggested by natural selection that only the mutants that exhibit optimal fitness and fit the environment survive. 
In computational algorithms, altering AA types of a wild-type protein can be viewed as adding corruptions to the node features of the protein graph, and recovering the perturbed graph becomes a remedy for identifying mutants with the best fitness. We thus model the protein mutation effect prediction as a denoising problem.

\begin{figure}[t]
    \centering
    \includegraphics[trim={0.7cm 2cm 7cm 3cm},clip, width=\textwidth]{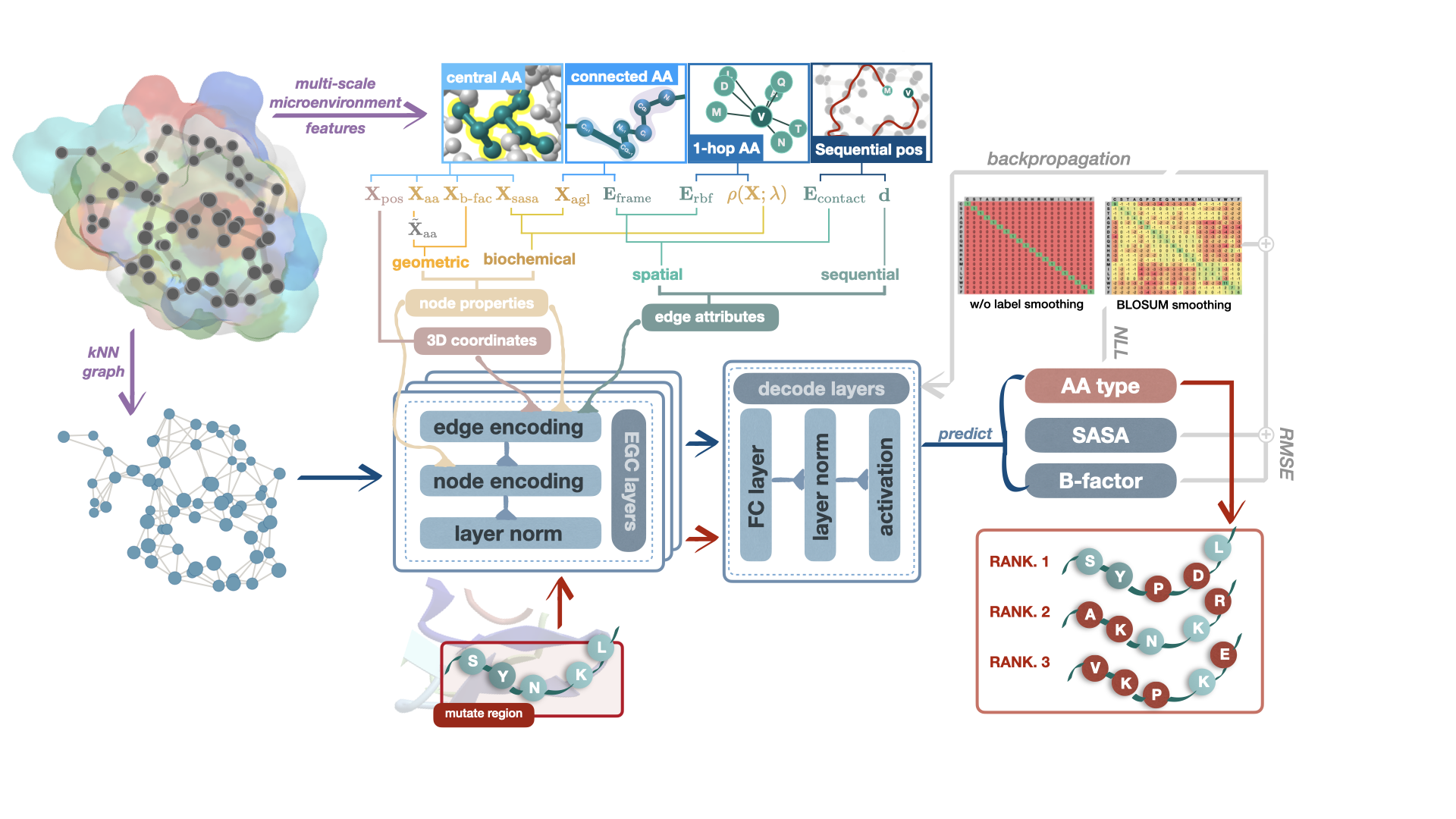}
    \caption{An illustration of the proposed method for variant effect predictions. \textcolor{purple!60!blue!80!black!80!}{(1) \textbf{Data Preparation}.} An input graph is represented by a $k$NN graph with node and edge attributes of the corresponding AAs extracted from multiple scales, as well as nodes' 3D positions. 
    \textcolor{cyan!50!black!70!}{(2) \textbf{Model Training}.} Wild-type protein graphs with noisy input attributes are fed to EGC layers for rotation and translation equivariant node embedding in the generic protein space. 
    \textcolor{red!60!black!70!}{(3) \textbf{Variant Effect Predictions}.} The fitness score of a mutant is derived from the joint distribution of the modified AA types on associated nodes. Updating the preliminary model with additional mutational samples can fit the specified protein and protein functionality better.}
    \label{fig:modelArchitecture}
\end{figure}

We provide a rich 3D spatial representation of the folded protein by a \emph{protein graph}, where each AA corresponds to a node in the graph. The node features encode important information such as AA types, spatial coordinates of C$\alpha$, and C-N angles between neighboring AAs. The protein graph inputs are then processed by equivariant graph neural networks (EGNNs) \citep{satorras2021n} in order to extract and utilize their geometric features in a rotationally-invariant manner. EGNNs provide a robust representation of AAs which are oriented differently depending on their location within the protein. It avoids costly data augmentation and leads to better performance in predicting mutations.

Our proposed model is lightweight since it begins with encoding proteins' structural and biochemical properties using a few layers of graph convolutions. The training precludes multiple sequence alignment (MSA) \citep{riesselman2018deep,frazer2021disease,rao2021msa} and protein language models \citep{ahmed2021protbert,rives2021biological,nijkamp2022progen2,brandes2022proteinbert}, both of which require substantial computational resources, such as hundreds of GPU cards, and massive data mining from millions of proteins. The high demand for computing resources hinders model revisions for the former approach, while the latter requires evolutionary properties of the protein family and considerable amounts of high-quality protein data for effective training.

Furthermore, our method circumvents the assumption of independent mutations in predicting the fitness of higher-order mutations by considering the joint distribution of AAs across the entire protein sequence. Traditional approaches that arrange autoregressive inference processes to find the conditional score for individual mutations at a single site are not only time-consuming for lengthy proteins but also based on an incorrect assumption that mutations on different sites occur sequentially or independently. The trivial assumption ignores epistatic effects among different sites, which are considered crucial factors in finding favorable high-order mutations, thus hindering directed evolution \citep{sato2019protein,liu2022rotamer,hsu2022esmif1,notin2022tranception}.

To summarize, the proposed lightweight equivariant graph neural network (\textsc{LGN}) has distinct advantages for predicting variant effects from three perspectives. First and foremost, it enables instant and highly reliable deep mutational effect inference that is specific to the protein and functionality. Secondly, the model is capable of generalizing to unseen proteins, making it a useful tool for recommending directional evolution strategies. Furthermore, \textsc{LGN} circumvents the independent-mutation assumptions and instead incorporates epistatic effects by utilizing the joint distribution of all variations. Empirically, we test \textsc{LGN} on $19$ proteins of up to 28-site mutations. On average, our model reaches $0.841$ Spearman's correlation in deep mutational effect predictions. It has outperformed the other supervised model \textsc{ECNet} \cite{luo2021ecnet} by $4.7\%$, and the rest SOTA self-supervised baseline models (\eg \textsc{DeepSequence} \citep{riesselman2018deep} and \textsc{ESM-IF1} \citep{hsu2022esmif1}) by $83.3\%$.

\begin{figure}[!tp]
    \centering
    \includegraphics[width=\textwidth]{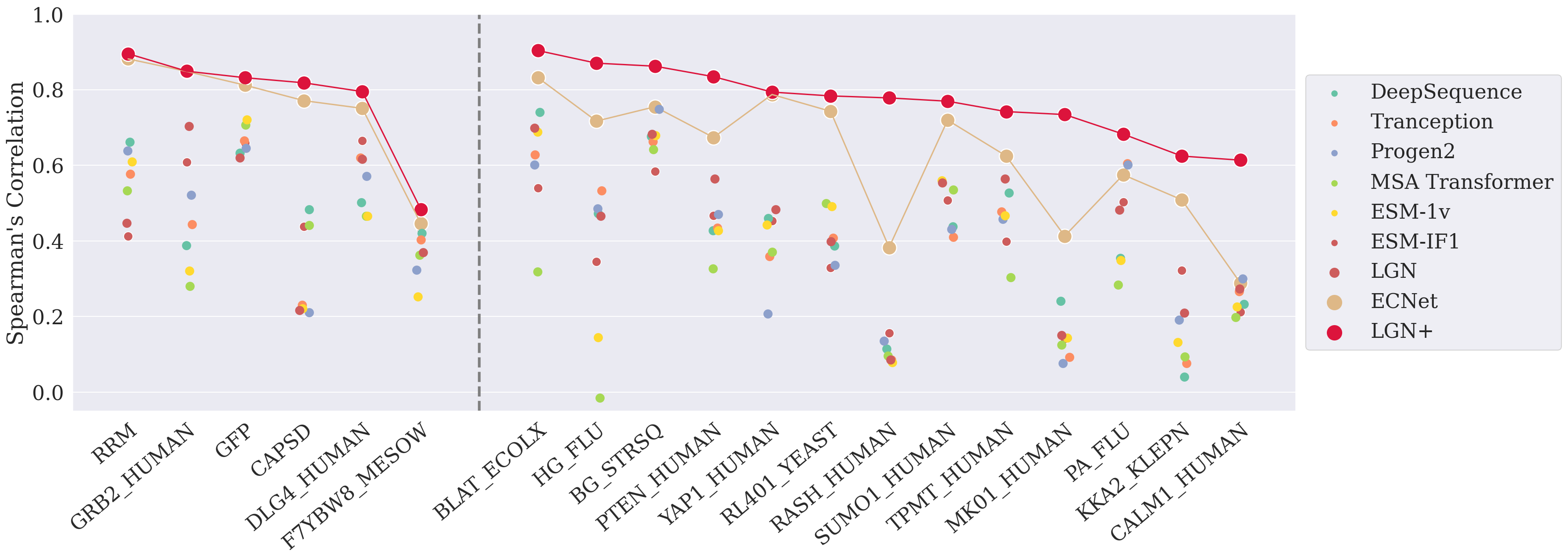}
    \caption{Per task performance on the variant effect prediction with models trained from zero-shot and supervised learning tasks. Colored scatters indicate Spearman's correlation coefficients on the corresponding protein achieved by different models. The left $6$ proteins contain higher-order mutations, and the right $13$ proteins record shallow mutants. }
    \label{fig:scatter:Nsite}
\end{figure}

\section{Results}
\label{sec:results}
The ability of \textsc{LGN} to predict variant effects is evaluated with deep mutational scanning (DMS) assays \cite{fowler2014deep}, which provides a systematic survey of the mutational landscape of proteins from wet-lab tests and is commonly used to benchmark computational predictors' effectiveness for evaluating mutations. Fig.~\ref{fig:modelArchitecture} displays the overall workflow of our model, where a protein graph with attributed nodes and edges (see construction rules in Section~\ref{sec:proteinGraph}) is inputted into equivariant graph convolutional layers to obtain appropriate node embeddings, which will be decoded for label prediction.

\subsection{Fitness of Deep Mutations Prediction}
Our model predicts the fitness scores following two different routes, depending on whether protein-specific true scores are available.  If there is no access to additional mutations, our model directly provides a preliminary assessment based on observations from nature and takes the log-odds-ratio from the predicted probabilistic distribution of the mutations. When a small set of mutations is available, the model will first be fine-tuned to predict accurate fitness scores of the underlying protein. The two models, without and with fine-tuning steps, are named \textsc{LGN} and \textsc{LGN+}, respectively.
The fitness score provides an overall assessment of the measurable characteristics of a protein in relation to specified mutations, such as enzyme function, growth rate, peptide binding, viral replication, and protein stability. A higher fitness score indicates that the mutant benefits from certain adjustments of associated sidechain types.

\begin{figure}[!tp]
    \includegraphics[width=\textwidth]{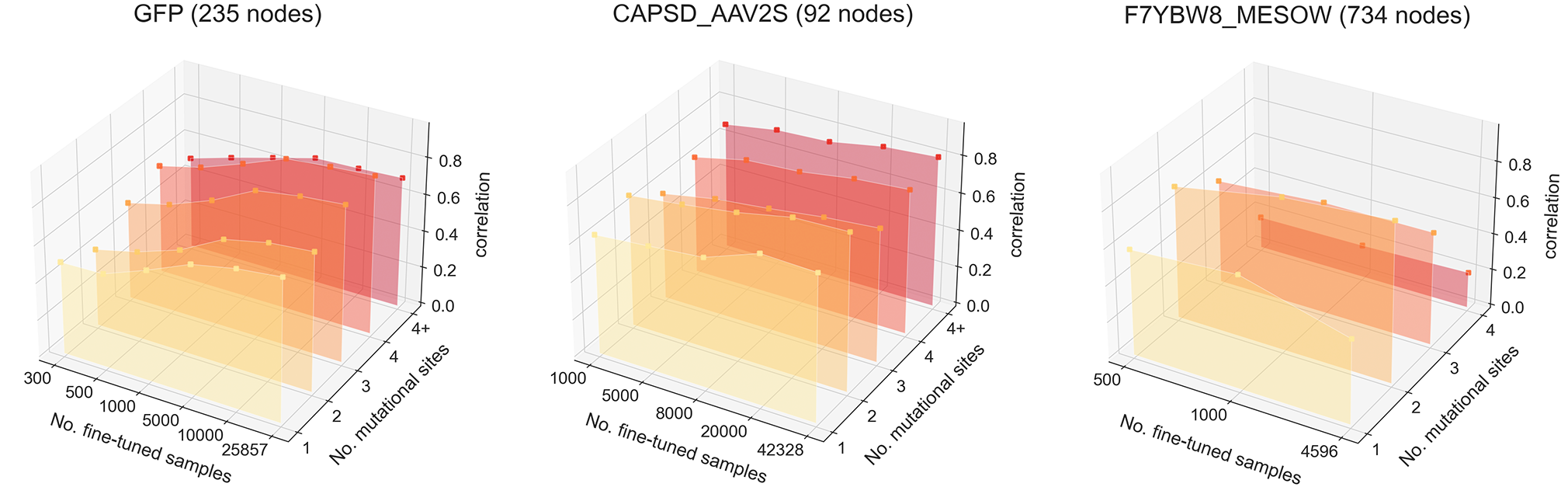}
    \caption{A small fraction of wet-lab test results are sufficient for greatly boosting LGN's prediction performance. }
    \label{fig:3dline:xsite}
\end{figure}
\begin{figure}[!t]
    \includegraphics[width=\textwidth]{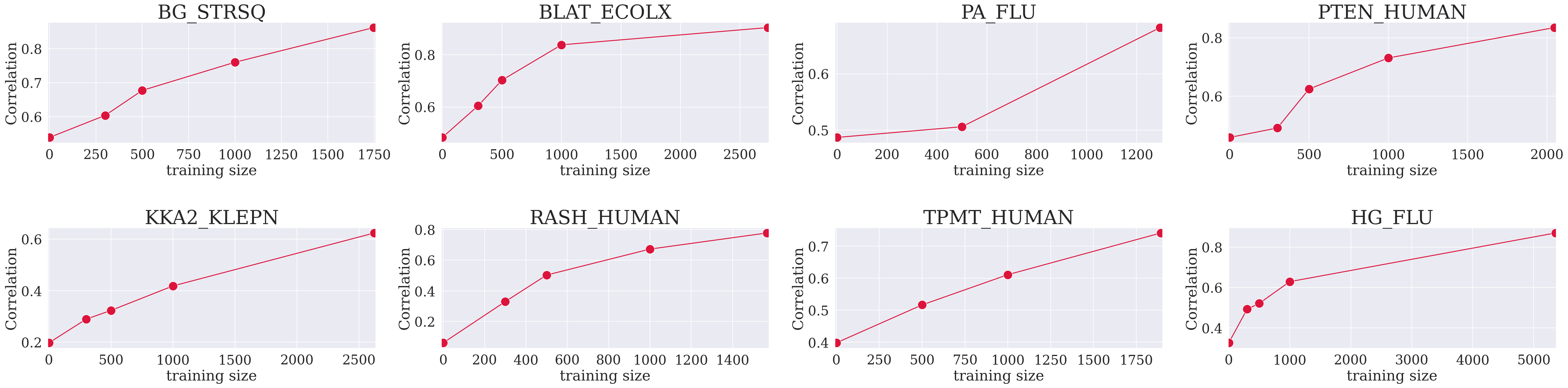}
    \caption{The Spearman's Correlation of fitness prediction on single-site mutations with different numbers of training samples.}
    \label{fig:line:1site}
\end{figure}

The comparison of LGN's performance is made with seven state-of-the-art baseline models on $299,120$ mutants from $19$ DMS experiments that cover $1$-site to $28$-site mutation scores in literature, where $13$ of them only mutate on single sites, and the rest $6$ of them test DMS assays on both single site and higher-order sites. To evaluate the reliability of variant effect predictions, we calculate protein-wise Spearman's correlation coefficient between the computational and experimental scores. The results are visualized in Fig.~\ref{fig:scatter:Nsite}. The preliminary LGN achieves comparable performance among other competitors, and minimum efforts advance LGN+ (denoted by the threaded crimson bullets~\textcolor[RGB]{209,0,47}{$\bullet$}) to substantially lead other competitors on different proteins. The enhanced LGN+ not only requires little computational resources but also consumes limited scored mutational samples to significantly improve the prediction performance. The number of fine-tuned samples for promising performance is investigated in Fig.~\ref{fig:3dline:xsite} and Fig.~\ref{fig:line:1site} for multiple-sites and single-site mutational predictions. It turns out that a few hundred can boost the correlation score, and up to thousands of mutational scores are sufficient for achieving a prominent correlation score even in the case of deep mutates. 


\begin{figure}[!tp]
    \includegraphics[width=\textwidth]{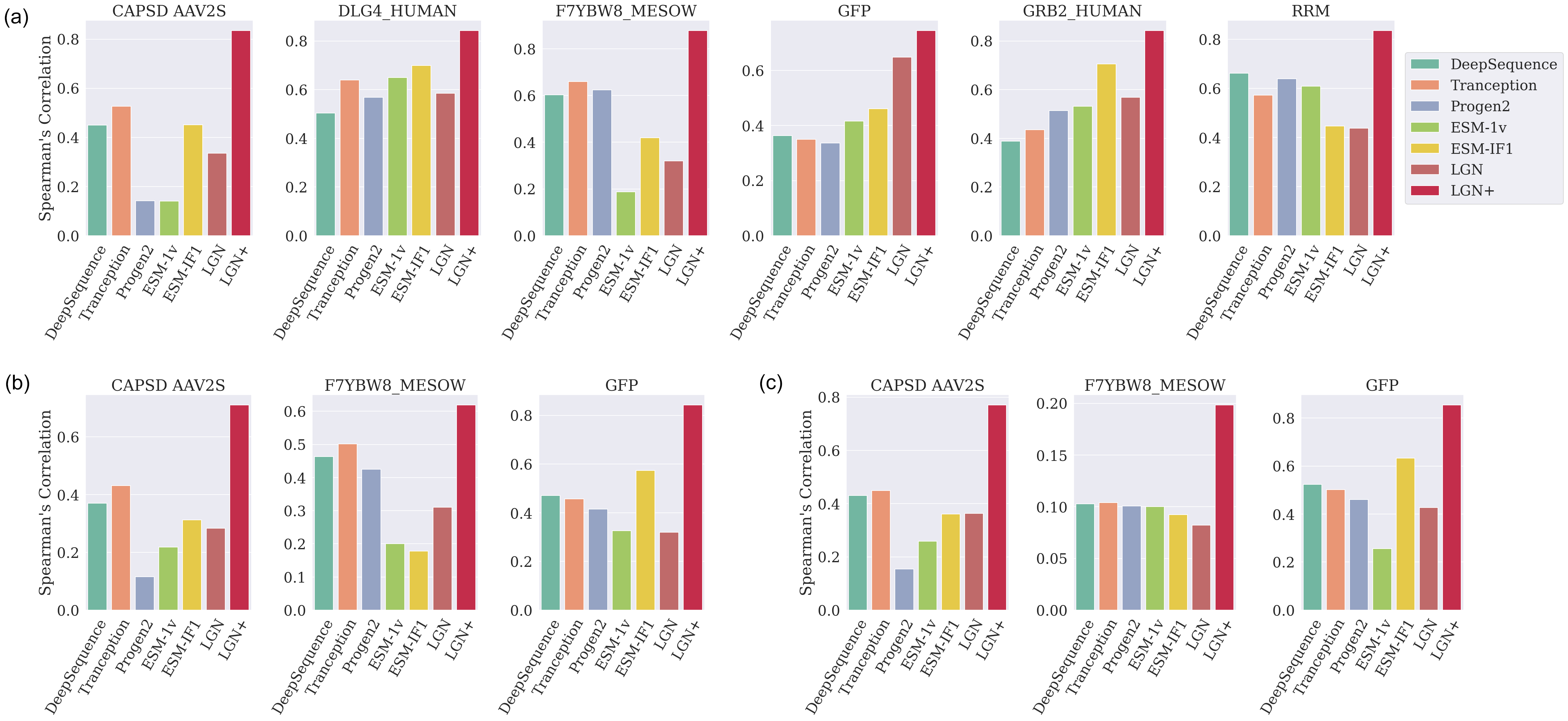}
    \caption{The Spearman's Correlation of fitness prediction on deep mutations. The three subplots (a), (b), and (c) are with respect to 2, 3, and 4-order mutations.}
    \label{fig:bar:xsite}
\end{figure}

\begin{table}[!tp]
\caption{Spearman's correlation and True positive rate (TPR) comparison of \textsc{ECNet} and LGN+ on the $13$ proteins of single-site mutations.}
\begin{center}
\label{tab:single_recall}
\resizebox{0.65\linewidth}{!}{%
    \begin{tabular}{lcccc}
    \toprule
    & \multicolumn{2}{c}{Spearman's Corr.} & \multicolumn{2}{c}{TPR (20\%)} \\ \cmidrule(lr){2-3}\cmidrule(lr){4-5}
    protein & \textsc{ECNet} & \textsc{LGN} & \textsc{ECNet} & \textsc{LGN} \\
    \midrule
    \textbf{BLAT\_ECOLX}  & $0.832$ & \bm{$0.903$} & $0.586$ & \bm{$0.728$} \\
    \textbf{HG\_FLU}      & $0.717$ & \bm{$0.870$} & $0.593$ & \bm{$0.759$} \\
    \textbf{BG\_STRSQ}    & $0.755$ & \bm{$0.862$} & $0.758$ & \bm{$0.805$} \\
    \textbf{PTEN\_HUMAN}  & $0.674$ & \bm{$0.834$} & $0.479$ & \bm{$0.699$} \\
    \textbf{YAP1\_HUMAN}  & $0.787$ & \bm{$0.794$} & $0.676$ & \bm{$0.743$} \\
    \textbf{RL401\_YEAST} & $0.743$ & \bm{$0.783$} & $0.421$ & \bm{$0.603$} \\
    \textbf{RASH\_HUMAN}  & $0.383$ & \bm{$0.778$} & $0.333$ & \bm{$0.557$} \\
    \textbf{SUMO1\_HUMAN} & $0.719$ & \bm{$0.769$} & $0.535$ & \bm{$0.688$} \\
    \textbf{TPMT\_HUMAN}  & $0.625$ & \bm{$0.742$} & $0.435$ & \bm{$0.611$} \\
    \textbf{MK01\_HUMAN}  & $0.413$ & \bm{$0.734$} & \bm{$0.337$} & $0.293$ \\
    \textbf{PA\_FLU}      & $0.575$ & \bm{$0.682$} & $0.306$ & \bm{$0.470$} \\
    \textbf{KKA2\_KLEPN}  & $0.509$ & \bm{$0.624$} & $0.588$ & \bm{$0.648$} \\
    \textbf{CALM1\_HUMAN} & $0.289$ & \bm{$0.614$} & $0.278$ & \bm{$0.607$} \\
    \bottomrule 
    \end{tabular}}
\end{center}
\end{table}

Fig.~\ref{fig:bar:xsite} details the evaluations on protein-specific and order-specific fitness prediction. 
Based on the available experimental records, the number of mutational sites is designated to 2-4 on the six proteins that explored deep mutational effects. It can be seen that LGN+ champions higher-order mutations by up to over 100\% improvement. The beneficial supervision of new experimental data is backed up by the lightweight model architecture of LGN, which is an advanced property that is exclusive to LGN. 
A comprehensive comparison of the training and inference cost is discussed in Section~\ref{sec:cost}. 

We also compare the two supervised learning models, \textsc{ECNet} and \textsc{LGN+}, by Spearman's correlation and true positive rate (for the top 20\%-ranked mutations) for single-site shallow mutations. Both methods take $10\%$ of the total records for training. According to the results reported in Table~\ref{tab:single_recall} and Fig.~\ref{fig:scatter:Nsite}, \textsc{LGN+} outperforms \textsc{ECNet} steadily and significantly on both metrics in predicting single-order mutational effects.
This could be explained by the fact that \textsc{ECNet}, as a completely supervised model, requires more training data to achieve satisfactory prediction performance. 
 Since generating mutational effect data through wet labs is usually expensive and time-consuming (due to the technical difficulties and the slow turnaround time of conducting experiments), it is desirable to tune a high-performance model with minimal supervision or with a small training set.

\subsection{Enhance Protein Embedding with Prior Knowledge}
To encourage the network to capture essential protein features, we integrate various types of prior knowledge into our pre-training procedure. Specifically, we incorporate perturbations on amino acid (AA) types to simulate potentially harmful mutations in wild-type proteins \cite{liu2022rotamer}. We also implement substitution matrices for noise generation and label smoothing to assist the generated mutations in accurately reflecting the natural variation.

\paragraph{AA Type Denoising} 
We refine the AA type of a node $\vx_{\text{aa}}$ to $\tilde{\vx}_{\text{aa}}$ with a Bernoulli noise, \ie 
\begin{equation}
\label{eq:noise_aa}
    \boldsymbol{\pi}(\tilde{\vx}_{\text{aa}} \mid \vx_{\text{aa}})=p\delta(\tilde{\vx}_{\text{aa}}-\vx_{\text{aa}})+(1-p)\boldsymbol{\Theta}(\pi_1,\pi_2,\dots,\pi_{20}),
\end{equation}
where the confidence level $p$ is a tunable parameter that controls the proportion of residues that are considered to be `noise-free'. It can also be determined by prior knowledge regarding the quality of wild-type proteins, \ie how frequently are mutations expected to happen in nature. For example, a value of $p=1$ indicates the maximum confidence in wild-type protein quality, which results in no perturbations to the input AA. 

The probability for the residue to become a particular type depends on the defined distribution $\boldsymbol{\Theta}$. A naive choice for $\boldsymbol{\Theta}$, the expected frequency distribution of AA types, is setting them to equal values: $\pi_1=\pi_2=...=\pi_{20}=0.05$, although it could be informed by prior knowledge based on molecular biology with the observed probability density of AA types in wild-type proteins. Substitution matrices can be another choice for defining pair-wise exchange probabilities while helping with robust representation learning through the use of label smoothing techniques. 

\begin{figure}[!t]
    \centering
    \includegraphics[width=\textwidth]{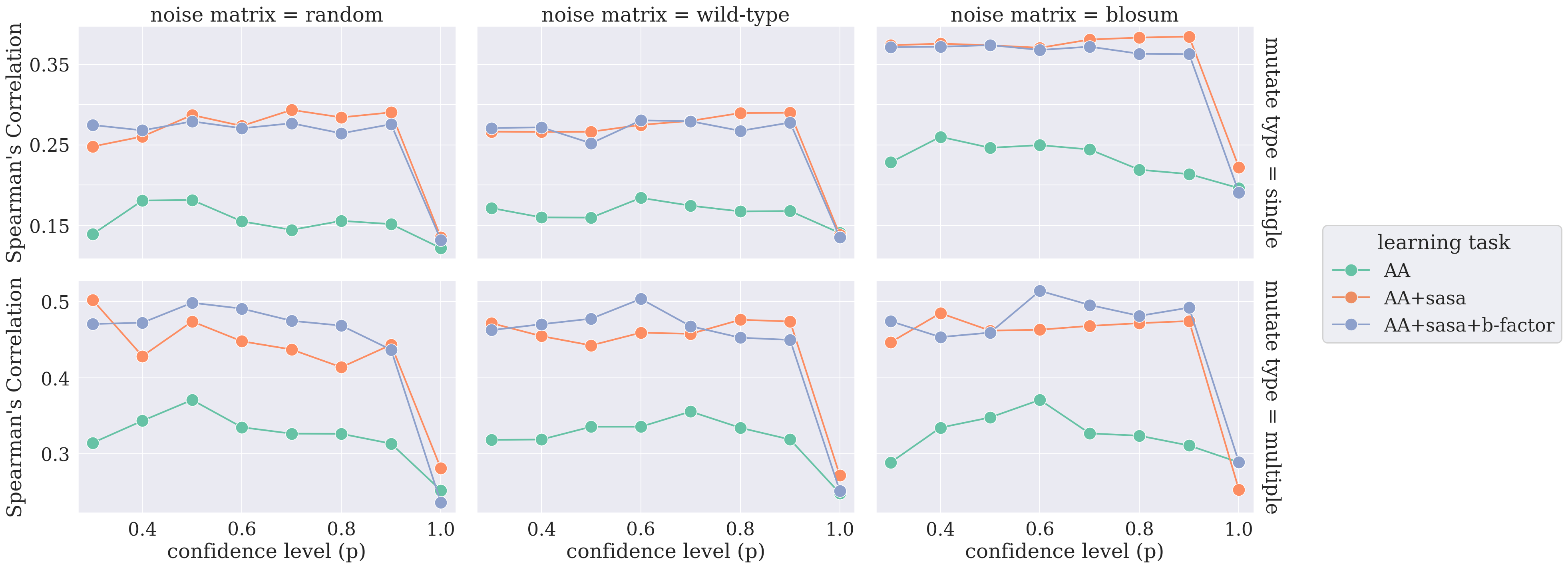}
    \caption{Average Performance with different $p$s on various AA noises (row) and mutational numbers. We use `mutate type=single' to indicate shallow 1-site mutations, and `mutate type=multiple' for deep mutations.}
    \label{fig:line:p+lambda}
\end{figure}

We evaluate the performance of the model with the above three types of exchange distributions $\boldsymbol{\Theta}$, as shown in Figure~\ref{fig:line:p+lambda}.
It examines the sensitivity of $p$'s choices by the average Spearman's correlation under different AA noise distributions, including random uniform distribution, wild-type AA distribution\footnote{Retrieved from the folded protein dataset by \textbf{AlphaFold 2} \citep{varadi2022alphafold} at \url{https://alphafold.ebi.ac.uk/}.}, and the substitution matrix. 
We are examining the general scenario where none of the mutants have undergone labeling or testing in wet labs. We have excluded extremely small values of $p$ to prevent drastic perturbation rates and to facilitate the search for an optimal $p\in\{0.3,0.4,\dots,0.9,1\}$. 
Different choices on $p$s are validated with three different learning tasks, which will be introduced shortly in Section~\ref{sec:multiTaskLearningResults}. 
A sharp decrease at $p=1$ certifies the effectiveness of introducing perturbations on AA types to the performance improvement of fitness prediction on both single-site and multiple-site mutations. 
Also, incorporating affinity learning tasks alongside AA sequence prediction facilitates the generation of more expressive node embeddings, as evidenced by the superior zero-shot fitness prediction performance. Notably, the prediction of SASA for each node using hidden protein graph embeddings yields significant enhancements compared to AA-only learning tasks. While including the additional B-factor predictor does not see significant advances on top of SASA prediction, it perfects the optimal model with better performance.
In general, assigning a moderate value to $p$ between $0.4$ and $0.7$ is suitable for different combinations of learning modules and noise distributions. Based on the overall performance, we have set the default value of p to 0.6 as the confidence level for our model.

\begin{figure}[!tp]
    \centering
    \includegraphics[width=\textwidth]{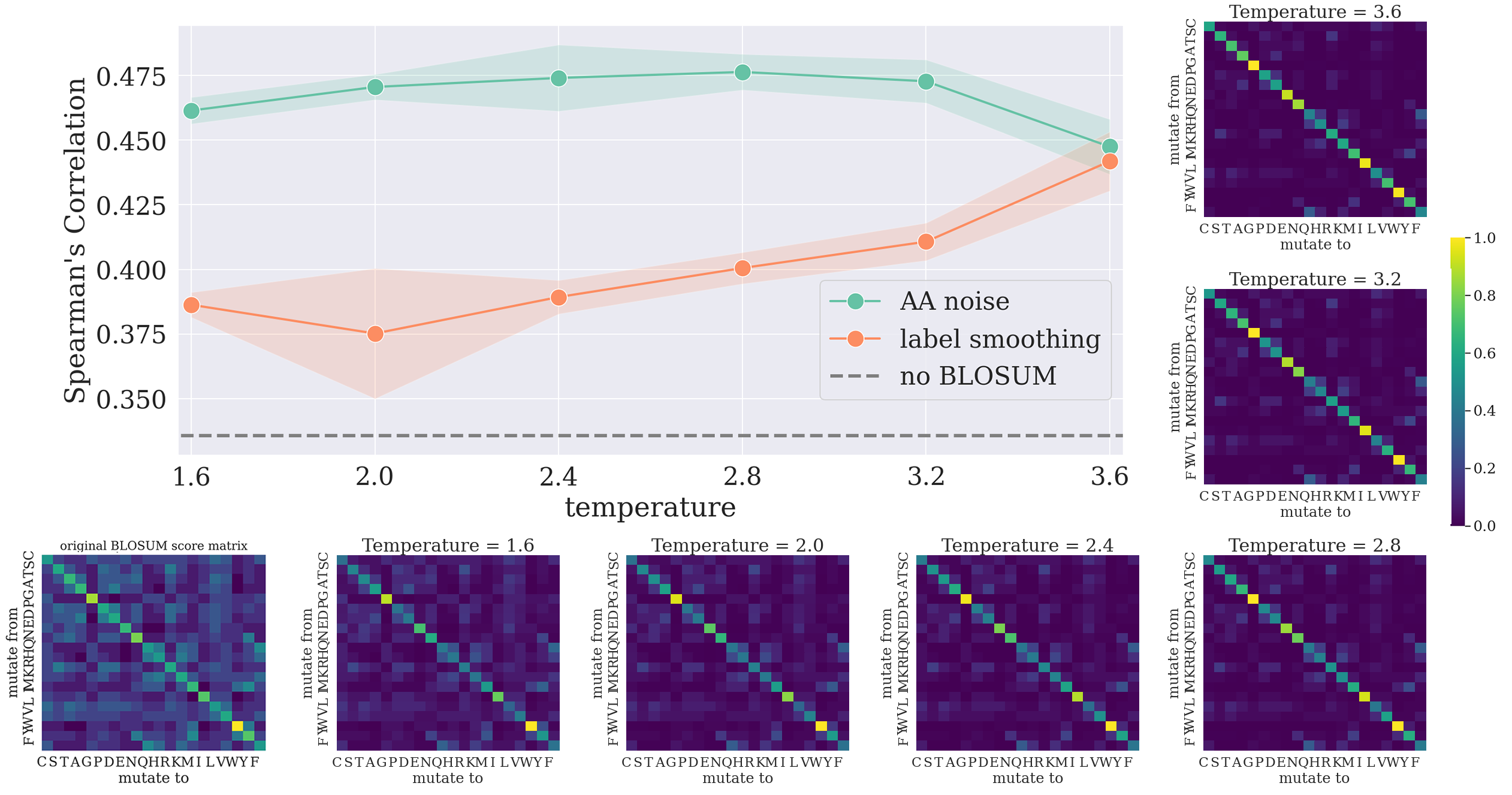}
    \caption{Average Spearman's correlation over $5$ repetitive runs on deep mutations with different temperatures on the \textsc{BLOSUM}62 matrix, where the matrix is used for generating AA noise (in orange) or label smoothing (blue). The normalized \textsc{BLOSUM}62 matrices as well as the original matrix are visualized at different temperature levels.}
    \label{fig:heat:blosum}
\end{figure}

\paragraph{Label Smoothing with Amino Acid Substitution Matrices}
\label{sec:labelsmoothing}
Protein sequence alignments provide important insights for understanding gene and protein functions. The similarity measurement of a protein sequence alignment reflects the favors of all possible exchanges of one AA over another. We employ \textsc{BLOSUM}62 substitution matrix \citep{henikoff1992amino} to account for the relative substitution frequencies and chemical similarity of AAs. The matrix is derived from the statistics for every conserved region of protein families in \textbf{BLOCKS} database. 
Given that AA sites are more likely to mutate to AA types with high similarity scores in the \textsc{BLOSUM}62 table, we used this matrix to modify our loss function with the label smoothing technique. Specifically, mutations to AA types with higher similarity scores accumulate smaller penalties compared to those with lower similarity scores. 

Fig.~\ref{fig:heat:blosum} demonstrates the modified \textsc{BLOSUM}62 matrix with different temperatures for defining the label smoothing and perturbation probability. The temperature $t$ is introduced to control the degree of dispersion towards off-diagonal regions, which transforms the substitution matrix $\mB$ to $\mB^{\prime}$ by $\mB^{\prime}=\sigma(\mB)^t$, where $\sigma(\cdot)$ is a non-linear operator, such as normalization. Intuitively, increasing $t$ pushes $\mB$ towards a diagonal matrix, and it is agnostic to a higher confidence level $p$ in wild-type noise, in the sense that both of them return more diagonal-gathered substitution matrices. The line plot reports the average correlation (over $5$ repetitive runs) of variant effect predictions for the $6$ deep mutational proteins with $p=0.6$. 
Compared to the baseline results of random AA noise and no label smoothing, applying the \textsc{BLOSUM}62 matrix to either AA noise or label smoothing improves the predictions. Overall, a higher temperature for the label smoothing matrix yields better overall performance, while a moderate temperature that produces more nuanced noise to AA types is preferred.

\subsection{Multitask Learning Strategy}
\label{sec:multiTaskLearningResults}
We utilize the multitask learning approach for the self-supervised learning module to enhance the microenvironment embedding of AAs and advance the expressivity of the hidden node representation. 

Initially, the model corrects the perturbed AA types and predicts the joint distribution of the types of all AAs, \ie $\vy_{aa}\in\mathbb{R}^{20}$. Concurrently, additional auxiliary tasks that predict $\vy_{\text{sasa}}$ and $\vy_{\text{B-fac}}$ introduce inductive biases to enhance the model's predictive capabilities. The former property, SASA, strongly influences AA type preferences, and the latter, B-factor, is associated with the conformations and mobility of the neighboring AAs. Both properties are closely connected to AAs and are essential in describing an underlying AA. Predicting the two features is thus beneficial, allowing the implicit encoding of the features without the risk of data leakage.

\begin{figure}[t]
    \centering
    \includegraphics[width=\textwidth]{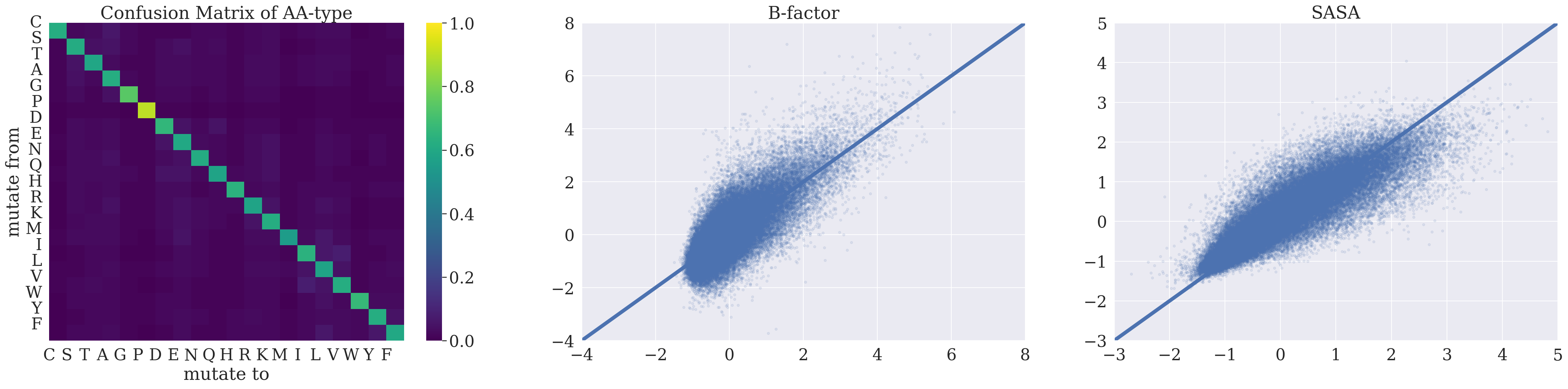}
    \caption{Confusion matrix of predicted AA types (left), and linear regression on the predicted SASA (middle) and B-factor (right).}
    \label{fig:heat_scatter:fitness}
\end{figure}

The efficacy of incorporating both auxiliary tasks has been previously examined in Fig.~\ref{fig:line:p+lambda}. The results demonstrate that including these supplementary targets significantly enhances performance, regardless of the selected noise distribution. 
Moreover, Fig.~\ref{fig:heat_scatter:fitness} indicates that all three tasks are well-learned during pre-training by reporting the predicted $\vy$s. Specifically, the confusion matrix of the predicted AA types with respect to the ground-truth AA types is visualized to assess the model's ability to recover from noisy sequences to the original sequence. 
The vast majority of predictions accumulating on the diagonal line indicate a high recovery rate with respect to AA types. For the two regression tasks, \ie SASA and B-factor predictions, we used linear regression to fit the true value against the predicted value, yielding estimated coefficients of $0.989$ and $1.008$, respectively, while both the corresponding $p$-values were found to be $<0.001$. Pearson's correlation coefficients for the predicted and true values of the two features are fairly high at $0.884$ and $0.791$, respectively.

\begin{figure}[t]
    \centering
    \includegraphics[width=0.7\textwidth]{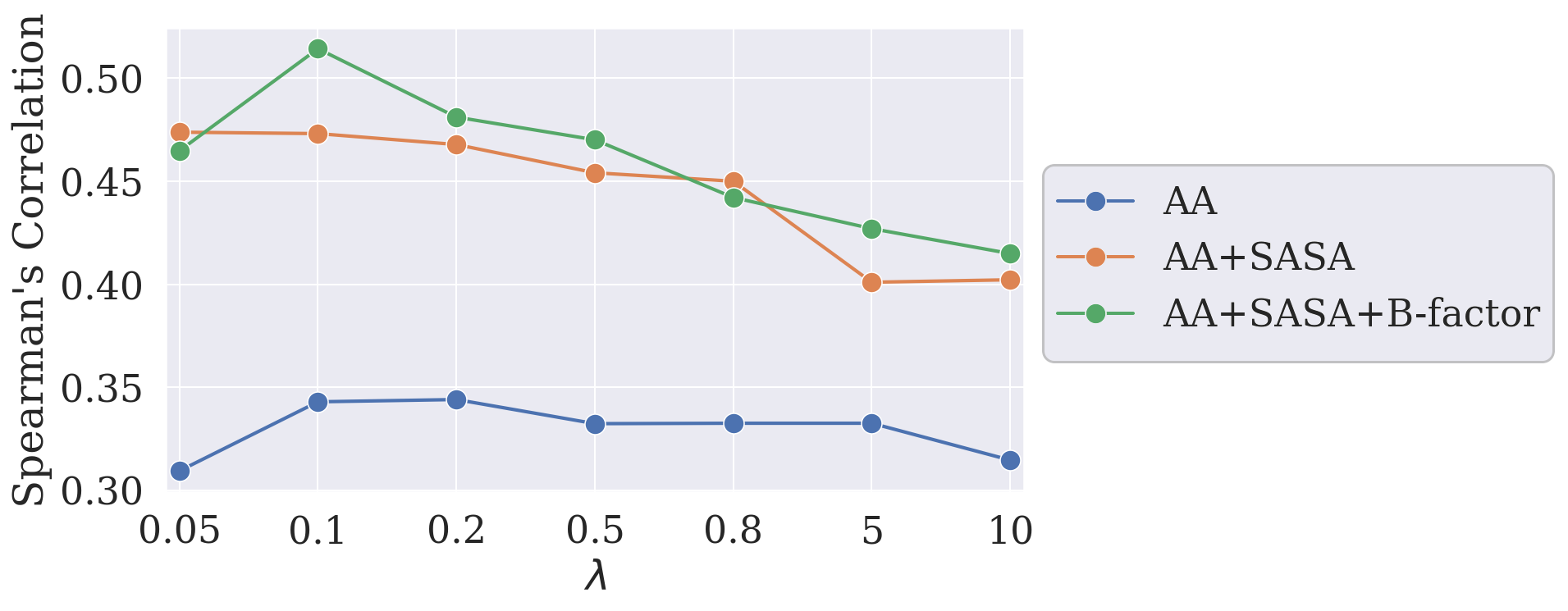}
    \caption{The influence of $\lambda$ on the loss of predicting $\vy_{\text{sasa}}$ and $\vy_{\text{B-fac}}$ to the overall correlation. }
    \label{fig:line:lambda}
\end{figure}

 It is crucial to balance the attention assigned to the three learning tasks and control their contribution to the overall prediction error $\gL_{\text{total}} = \gL_{\text{aa}}+\lambda_1\gL_{\text{sasa}}+\lambda_2\gL_{\text{B-fac}}$ with factors $\lambda_1, \lambda_2$. Here we investigate a wide range of the choices of $\lambda$s to the impact of model training. As both $\vy_{\text{sasa}}$ and $\vy_{\text{B-fac}}$ share a consistent value scale, we let $\lambda_1=\lambda_2\in\{0.05, 0.1, 0.2, 0.5, 0.8, 5, 10\}$. We conducted all experiments with wild-type noise and set $p=0.6$. The average Spearman's correlations, reported in Fig.~\ref{fig:line:lambda}, exhibit an overall decreasing trend, highlighting the importance of accurately predicting $\vy_{\text{aa}}$ as the primary objective for the model.
 The scores corresponding to $\lambda$ values from $0.05$ to $0.1$ indicate that including the two auxiliary tasks is necessary, particularly when making all three predictions simultaneously. Furthermore, the overall trend suggests that accurately predicting $\vy_{\text{aa}}$ remains the model's primary objective. Notably, there is a small peak in the scores between $\lambda=0.1$ and $\lambda=0.2$, implying that smaller values of $\lambda$ are generally preferable.

\subsection{Training Cost of Self-Supervised Models}
\label{sec:cost}
Introducing abundant prior domain knowledge not only let \textsc{LGN} achieve excellent performance in variant effects prediction tasks with interpretable designs, but also significantly reduced the computational resources required during both model training and inference. The former advantage of eased and faster network training is particularly favored for instant model optimization and modification on specific proteins or directed evolution targets. 

\begin{figure}[t]
    \centering
    \includegraphics[width=\textwidth]{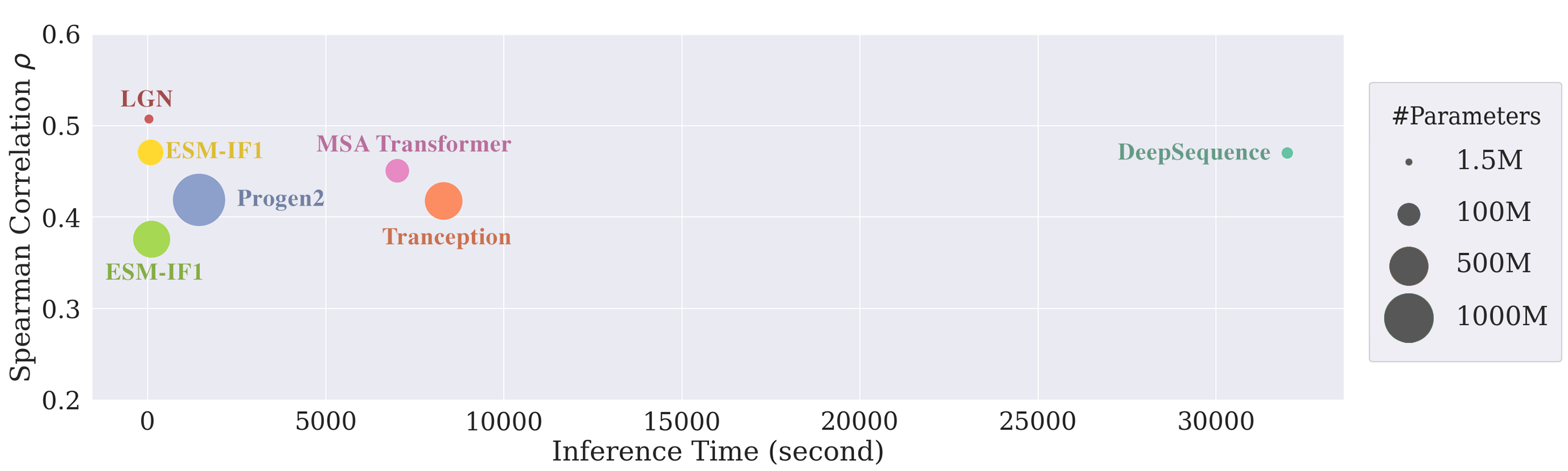}
    \caption{Comparison of Inference Efficiency on pre-trained models. The area of the ball indicates the number of network parameters of a model. Our model (in blue) can achieve SOTA performance ($y$-axis) with minimum inference time ($x$-axis) and $1\%$ number of parameters of the ESM.}
    \label{fig:bar:inferenceTime}
\end{figure}

Fig.~\ref{fig:bar:inferenceTime} and Table~\ref{tab:baselineComparison} deliver a direct comparison of our model against the baseline methods with respect to the model scale, inference time, and prediction performance. As the majority of models are pre-trained, we record the inference speed on a single NVIDIA GeForce RTX 3090. Note that the inference time varies with the length of the protein sequence and the size of the test set, we thus take an example protein \textbf{GFP} that consists of $236$ AA residues and has evaluation scores on over $50,000$ mutants for testing. Although all these \textit{in silico} computational costs are significantly lower than wet lab tests, we use the inference time to indicate the cost of forward propagation in one iteration, which is proportional to the model scale. In terms of \textsc{ECNet}, the running time also depends on the number of available training samples and the highest number of mutational orders. In the case of \textbf{GFP}, it takes $8,280$ seconds to finish the training procedure on around $5,200$ mutants (\ie $10\%$ of the total DMS samples).

\begin{table}[!tp]
\caption{Comparison of baseline zero-shot pre-trained models. The train and inference speed is tested on \textbf{GFP}.}
\begin{center}
\label{tab:baselineComparison}
\resizebox{\linewidth}{!}{
    \begin{tabular}{lccccccc}
    \toprule
    \textbf{model} & \textsc{DeepSequence} & \textsc{MSA Trans.}  & \textsc{ESM-1v} & \textsc{ESM-IF1} & \textsc{Tranception} & \textsc{ProGen2} & LGN (ours) \\
    \midrule
    input & sequence & sequence & sequence  & sequence+structure & sequence & sequence & structure \\
    MSA & $\checkmark$ & $\checkmark$ &&& $\checkmark$ &\\
    train on new protein & $\checkmark$ & &&& $\checkmark$ & \\
    training dataset & - & \textbf{Uniref50} &  \textbf{Uniref90} & \textbf{CATH+AF2} & \textbf{Uniref100} & \textbf{Uniref90}+\textbf{BFD30} & \textbf{CATH} \\
    & - & (2018-03) & & (2020-03) & && v4.3.0 \\
    training size (M) & - & $45$ & $98$ & $12$ & $249$ & $>1,000$ & $0.03$ \\
    max. input token & & $1,024$ & $1,024$ & $1,024$ & $1,280$ & $1,024$ & $2,687^{\dagger}$\\ 
    \midrule
    \# parameters (M) & $4.3$ & $100$ &  $650$ & $142$ & $700$ & $2,700$ & $1.5$ \\
    \# layers & $1,600$ & $12$  & - & $20$ & $36$ & $32$ & $6$\\
    \# head & - & $12$  & - & $8$ & $20$ & $32$ &  - \\
    \# hid. dim. & $100-2,000$ & -  & - & $512-2,048$ & - & - & $512$ \\
    speed (training day) & - & $13^{\dagger\dagger}$ & $6$ & $653$ & $\sim$100 & - & $0.17$ \\
    resource (train) & - & 128$\times$V100$^{\dagger\dagger}$ & 64$\times$V100 & 32$\times$V100 & 64$\times$A100 & ?$\times$TPU-v3 & 1$\times$3090 \\
    \midrule
    preparing speed (sec) & $6,360+25,020$ & $6,360$ & - & - & $6,360$ & - & - \\
    inference speed (sec) & $608$ & $927$ & $75$ & $102$ & $1,920$ & $1,440$ & $25$  \\
    \bottomrule  \\[-2.5mm]
    \multicolumn{8}{l}{$\dagger$ The $2,687$ input token length only refers to the maximum protein length we used during training, rather than its maximum capacity.} \\ 
    \multicolumn{8}{l}{$\dagger\dagger$ The training speed and required resources are retrieved from \cite{meier2021esm1v}.} \\
    \end{tabular}}
\end{center}
\end{table}

\section{Methods}
\label{sec:method}
When designing an expressive geometric deep learning model for predicting DMS assays on proteins, two principles should be carefully considered. Firstly, based on the laws of physics, the atomic dynamics of proteins remain unchanged, regardless of their translation or rotation from one position to another.\citep{han2022geometrically}. Therefore, the inductive bias of symmetry should be incorporated into the design of protein structure-based models. This ensures that the spatial relationship of AAs \citep{torng20173d, sato2019protein} or geometric equivariance \citep{ganea2021independent,stark2022equibind} are respected. Second, a protein encoder that is both expressive and general is required to balance the conflicts between scarce mutational records and the substantial resources required to train representation learning models. As with many studies \citep{ahmed2021protbert,zhang2022protein,hsu2022esmif1,castro2022transformer}, this work starts from a self-supervised method for discovering expressive representation for the rational protein space. On the contrary to related works in literature, we did not employ multiple sequence alignment (MSA) \cite{riesselman2018deep,jumper2021highly,rao2021msa,wang2022lm}. This is because not all proteins are alignable (such as CDR regions of antibody variable domains \citep{shin2021protein}) and not all the alignments are deep enough to train models sufficiently large capable of learning the complex interactions between residues.
Instead, we approach fast and robust modeling with proteomic knowledge and multitask learning strategies. When there are additional experimental records, the constructed model can be revised in the later stage.

\begin{figure}[!t]
    \centering
    \includegraphics[trim={2cm 7cm 2cm 1.8cm},clip, width=\textwidth]{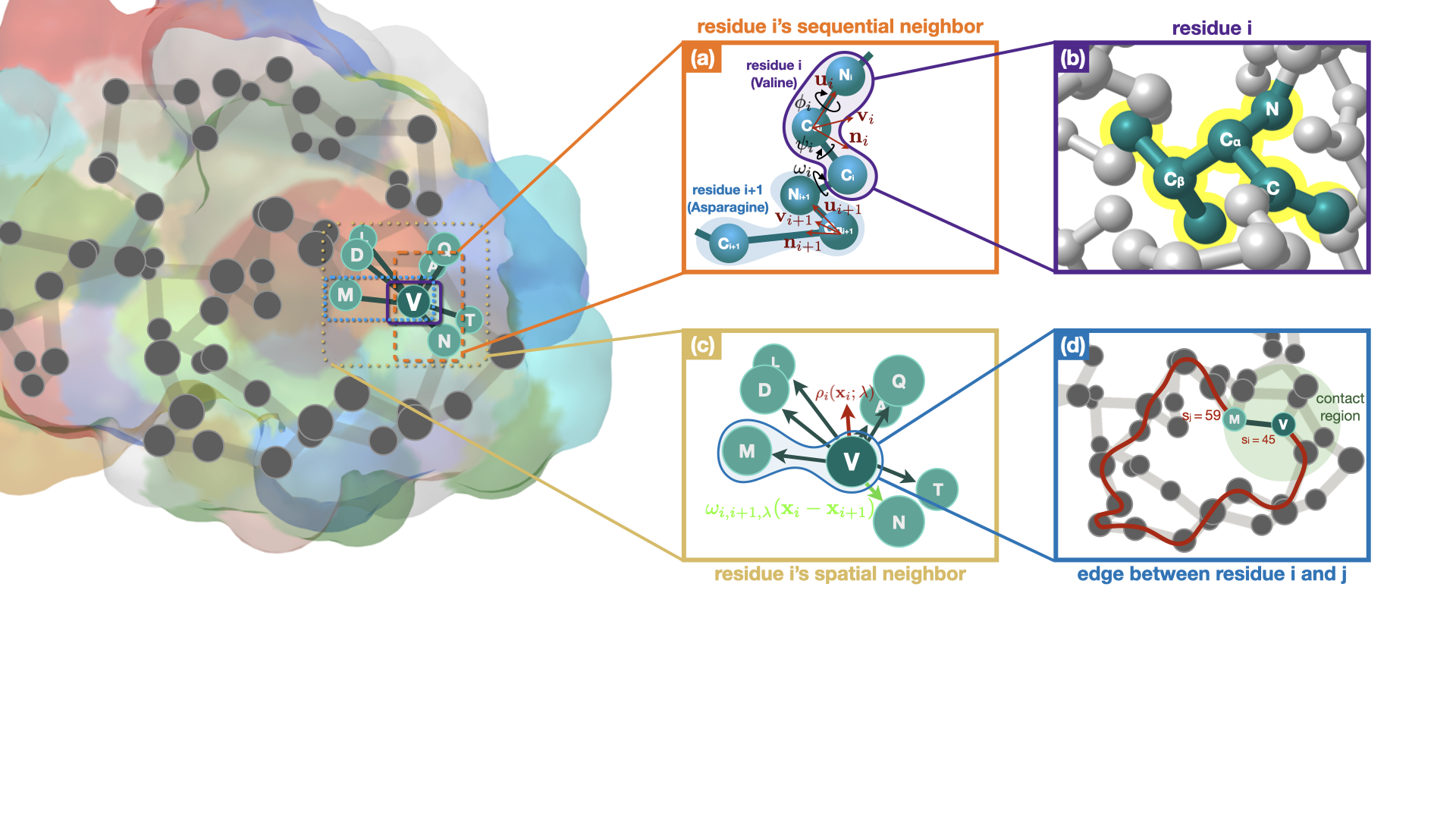}
    \caption{Illustrative node and edge features of a protein graph (PDB ID: 1KDF). For a particular node V45, \ie Valine at position 45: (a) the dihedral angle and local frames are constructed by the positions of the heavy atoms of V45 and N46 (the next residue on the sequence), where the local frames are also used for generating neighboring projections for edge features; (b) the AA type, B-factor, and SASA are generated from the raw \textit{.pdb} input; (c) the surface-aware features describe the weighted sum force of 1-hop neighbors, and the scale of this force implies whether it's an interior or surface-closed AA; (d) sequential relative positions and contact indicators are encoded in the edge attributes of connected nodes. }
    \label{fig:proteinGraphFeature}
\end{figure}

\subsection{Graph Representation of Folded Proteins}
\label{sec:proteinGraph}
The geometry of proteins suggests higher-level structures and topological relationships, which are vital to protein functionality. For a given protein, we create a $k$-nearest neighbor ($k$NN)  graph $\gG=(\mX,\mA,\mE)$ to describe its 3D structure and molecular properties. Here each node represents an AA with $\mX\in\mathbb{R}^{36}$ node attributes describing $21$ biochemical and $15$ geometric properties of AAs. The edge connections are formulated by a symmetric adjacency matrix $\mA$ with the $k$NN-graph to capture the nodes' microenvironment, \ie each node is connected to up to $k$ other nodes in the graph that has the smallest Euclidean distance over other nodes, and the distance is smaller than a certain cutoff (\eg $30\AA$). Consequently, if node $i$ and $j$ are connected to each other, we have $\mA_{ij}=\mA_{ji}\neq0$ with edge features $\mE\in\mathbb{R}^{93}$ defined on them. We now introduce the node and edge features.

The \emph{biochemical} node features include $20$ one-hot encoded AA types ($\mX_{\text{aa}}$) and a scalar value, \ie the standardized crystallographic B-factor, that identifies the rigidity, flexibility, and internal motion of each residue. Note that the raw B-factor is sensitive to the experimental environment and proteins in our dataset are measured by different laboratories, therefore we decide to fix the measurement bias by taking standardized B-factors along each protein. Specifically, we standardize the raw B-factor values with AA-wise mean and standard deviation. 

Regarding the \emph{geometric} node attributes, we include solvent-accessible surface area (SASA), normalized surface-aware node features, dihedral angles of backbone atoms, and 3D positions. SASA measures the level of exposure of an AA to solvent in a protein, which provides an important indicator of active sites of proteins to locate whether a residue is on the surface. The $5$ mean force features implement a non-linear projection to the weighted average distance of a residue to their one-hop neighbors $i^{\prime} \in \mathcal{N}_i$, \ie 
\begin{equation*}
      \rho\left(\mathbf{x}_i ; \lambda\right)=\frac{\left\|\sum_{i^{\prime} \in \mathcal{N}_i} w_{i, i^{\prime}, \lambda}\left(\mX_{\text{pos},i}-\mX_{\text{pos},i^{\prime}}\right)\right\|}{\sum_{i^{\prime} \in \mathcal{N}_i} w_{i, i^{\prime}, \lambda}\left\|\mX_{\text{pos},i}-\mX_{\text{pos},i^{\prime}}\right\|}, 
\end{equation*}
where the weights are defined by $w_{i, i^{\prime}, \lambda}=\frac{\exp \left(-\left\|\mX_{\text{pos},i}-\mX_{\text{pos},i^{\prime}}\right\|^2 / \lambda\right)}{\sum_{i^{\prime} \in \mathcal{N}_i} \exp \left(-\left\|\mX_{\text{pos},i}-\mX_{\text{pos},i^{\prime}}\right\|^2 / \lambda\right)}$ with $\lambda\in\{1,2,5,10,30\}$. These features describe whether the node is on the surface of the protein. 
A surface-closed AA with neighbors from a narrower range leads to larger feature values and stronger mean forces \cite{ganea2021independent}.
The $\mX_{\text{pos},i}\in\mathbb{R}^3$ denotes the \emph{3D coordinates} of the $i$th residue, which is represented by the position of $\alpha$-carbon. Moreover, the spatial conformation of the AA in the protein chain is measured by $\mX_{\text{agl},i}\in\mathbb{R}^6$, which contains the trigonometric values of dihedral angles $\{\sin, \cos\} \circ \{\phi_i, \psi_i,\omega_i\}$ of the backbone atom positions. The three dihedral angles $\phi_i, \psi_i$, and $\omega_i$ describe the torsion angle between the heavy atoms $N_i-C\alpha_{i}, C\alpha_{i}-C_i$, and $C_i-N_{i+1}$. The last nodes of AA sequences are removed to avoid inaccessible angles. 

The edge attributes $\mE\in\mathbb{R}^{93}$ feature the connected edges in the graph, including $15$ high-dimensional distances, $12$ relative spatial positions, and $66$ relative sequential distances. For two connected residues $i$ and $j$, the distance between them is projected by Gaussian radial basis functions (RBF), \ie
\begin{equation*}
      \mE_{r}^{\rm rbf}(\vx_i,\vx_j)= \exp\left\{\frac{\|\vx_j-\vx_i\|^2}{2\sigma^2_r}\right\}, \quad r=1,2,\dots,R.
\end{equation*}
A total number of $15$ distinct distance-based features are created on the edge with the scale parameter $\sigma_r=\{1.5^k\mid k=0,1,2,\dots,14\}$. From the corresponding residues' heavy atoms positions, the two local frames define $12$ relative positions \cite{ganea2021independent} of node $j$ with respect to node $i$. They represent local fine-grained relations between AAs and the rigid property of how the two residues interact with each other. Finally, the residues' sequential relationship is encoded with $66$ binary features by their relative position $d_{i,j} = \vert s_i - s_j\vert$, where $s_i$ and $s_j$ are the absolute positions of the two nodes in the AA chain. For instance, if $v_i$ is the first AA in the sequence and $v_j$ is the fifth AA, we have $s_i=1$ and $s_j=5$. We set a cutoff at $64$, \ie $d_{i,j} = \min(\vert s_i - s_j\vert, 65)$ based on the fact that the locally connected nodes (by the $k$NN defined edges) merely have their positional distance over $64$ and transform this distance feature with one-hot encoding \cite{liu2022rotamer}. In addition, we define a binary contact signal $\mE_{\text{contact}}$ \cite{ingraham2019generative} to indicate whether two residues contact in the space, \ie the Euclidean distance $\|C\alpha_i-C\alpha_j\|<8$. 

\subsection{Equivariant Protein Graph Convolution}
Bio-molecules such as proteins and chemical compounds are structured in the 3-dimensional space, and it is vital for the model to predict the same binding complex no matter how the input proteins are positioned and oriented. Instead of practicing expensive data augmentation strategies, we construct SE(3)-equivariant neural layers \cite{satorras2021n} for graph embedding. At the $l$th layer, an Equivariant Graph Convolution (\textsc{EGC}) inputs a set of $n$ hidden node properties embedding $\mH^{l}=\left\{\vh_1^{l}, \dots, \vh_n^{l}\right\}$ as well as the node coordinate embeddings $\mX_{\text{pos}}^{l}=\left\{\vx_1^{l}, \dots, \vx_n^{l}\right\}$ for a graph of $n$ nodes. The attributed edges are denoted as $\mE=\{\dots,\ve_{i j},\dots\}$. The target of an \textsc{EGC} layer is to output a transformation on the node feature embedding $\mH^{l+1}$ and coordinate embedding $\mX_{\text{pos}}^{l+1}$. Concisely: $\mH^{l+1}, \mX^{l+1}_{\text{pos}}=\operatorname{EGC}\left[\mH^{l}, \mX^{l}_{\text{pos}}, \mE\right]$, \ie
\begin{equation}
\label{eq:egnn}
    \begin{aligned}
    \vm_{ij} &=\phi_{e}\left(\mathbf{h}_{i}^{l}, \mathbf{h}_{j}^{l},\left\|\mathbf{x}_{i}^{l}-\mathbf{x}_{j}^{l}\right\|^{2}, \ve_{ij}\right) \\
    \vx_{i}^{l+1} &=\mathbf{x}_{i}^{l}+\frac1{n}\sum_{j \neq i}\left(\mathbf{x}_{i}^{l}-\mathbf{x}_{j}^{l}\right) \phi_{x}\left(\mathbf{m}_{i j}\right) \\
    \vh_{i}^{l+1} &=\phi_{h}\big(\mathbf{h}_{i}^{l}, \sum_{j \neq i} \mathbf{m}_{i j}\big),
\end{aligned}
\end{equation}
where $\phi_e, \phi_h$ are respectively the edge and node propagation operations, such as multi-layer perceptrons (MLPs). The $\phi_x$ is an additional operation that projects the vector embedding $\vm_{ij}$ to a scalar value. 
An EGC layer first aggregates representations of node pairs with their edge attributes and the Euclidean distance between the nodes. Next, the nodes' 3D positions for the next layer are updated with the projected propagated embedding ($\phi_x(\vm_{ij})$) as well as the differences in the coordinates of neighboring nodes within the 1-hop range. In the final third step, the hidden embedding for the node $i$ is updated by a conventional message passing of node $i$ and its 1-hop neighbors' hidden embedding from the previous steps. The EGC layer preserves equivariance to rotations and translations on the set of 3D node coordinates $\mX_{\text{pos}}$, while simultaneously performing invariance to permutations on the nodes set identical to any other GNNs.

\subsection{Multitask Learning for Model Pre-training}
\label{sec:multitaskLearningMethod}
The excessive cost in the laboratory results in scarce mutational scanning data, especially deep mutational results. It is thus favorable to first pre-train a zero-shot protein prediction model that can discover the generic protein space. The general-purpose protein model (where the model can be applied to various proteins and goals) is expected to learn essential information from self-supervision so that it can be applied directly to a variety of unseen new tasks without further specialization. Moreover, the pre-trained model benefits follow-up learning procedures by consuming fewer training samples and time to analyze a specific dataset, as it learns the generic patterns from large protein datasets. The consequent fine-tuned models frequently lead to better performance with improved generalization.

After the \textsc{EGC} layers extract rotation and translation equivariant node representations $\mH^{\text{out}}$ on individual graphs, the hidden representation is sent to fully-connected layers to establish node properties prediction. To approach meaningful and robust representations for the AAs' local environment, we require the output embedding $\mH^{\text{out}}$ of the nodes to accurately predict several key properties, including AA type classification, and SASA and B-factor value prediction. 
To be clear, the ground-truth SASA and B-factor will be excluded from the input feature when they become predictors. This is different from predicting AA types, where noisy AA labels are always provided in the input. The strategy is implemented by multitask learning, the total loss of which is given by
\begin{equation}
\label{eq:lossFunc}
    \gL_{\text{total}} = \gL_{\text{aa}}+\lambda_1\gL_{\text{sasa}}+\lambda_2\gL_{\text{B-fac}},
\end{equation}
where $\lambda_1, \lambda_2$ are tunable hyper-parameters to balance different losses on auxiliary regression tasks. Both $\gL_{\text{sasa}}$ and $\gL_{\text{B-fac}}$ are measured by the mean squared error (MSE). For AA type classification, $\gL_{\text{aa}}$ is measured by cross-entropy with label smoothing technique \citep{szegedy2016rethinking} to tolerant AA substitution among similar classes. The smoothed loss on an arbitrary node $i$ reads
\begin{equation*}
\begin{split}
    \gL_{\text{aa}} =& (1-\varepsilon) \Big[-\sum_{\vy=1}^{20} p(\vy_{\text{aa}} \mid \mX_i,\mE_i) \log q_{\theta}(\hat{\vy}_{\text{aa}}\mid \mX_i,\mE_i) \Big] \\
    &+ \varepsilon \Big[-\sum_{\vy=1}^{20} u(\vy_{\text{aa}}\mid \mX_i,\mE_i) \log q_{\theta}(\hat{\vy}_{\text{aa}}\mid \mX_i,\mE_i) \Big],
\end{split}
\end{equation*}
where $p(\vy_{\text{aa}}\mid \mX_i,\mE_i)$ denotes the ground-truth distribution of node $i$ to have a specific AA type, and $q_{\theta}(\hat{\vy}_{\text{aa}}\mid \mX_i,\mE_i)$ is the distribution of predicted labels following a softmax function. The tolerance factor $\varepsilon$ is a tunable hyper-parameter. In order to improve the generalization and respect the prior biological knowledge, we modify the ground truth label distribution $p(\vy_{\text{aa}}\mid\mX_i,\mE_i)$ from the hard one-hot encoding to $(1-\varepsilon) p(\vy_{\text{aa}}\mid\mX_i,\mE_i)+\varepsilon u(\vy_{\text{aa}}\mid\mX_i,\mE_i)$ when the predicted $\hat{\vy}_{\text{aa}}=\vy_{\text{aa}}$ and to $\varepsilon u(\vy_{\text{aa}}\mid\mX_i,\mE_i)$ otherwise. The distribution $u(\vy_{\text{aa}}\mid \mX_i)$ is approximated by the \textsc{BLOSUM}62 substitution matrix.

\subsection{Variant Effect Scoring}
The recovered protein sequence (and the corresponding predicted AA type distribution $\vy_{\text{aa}}\in\mathbb{R}^{20}$) by a pre-trained model to some extent reflects the rational appearance of a protein that is selected by nature. When an arbitrary protein with a cold-start (\ie no experimentally tested data at the beginning) is investigated, we follow \citep{meier2021esm1v,lu2022machine} and define \emph{log-odds-ratio} as a substitution of the fitness score that is obtained directly from probabilities. 
For $T$-site mutants ($T\geq 1$), the fitness score reads
\begin{equation}
\label{eq:logRatioScore}
    \sum_{t\in T} \log p(\vy_{\text{aa}}=\hat{\vy}_{\text{aa}}^{\text{mutate}}) - \log p(\vy_{\text{aa}}=\vy_{\text{aa}}^{\text{wild}}),
\end{equation}
where $\hat{\vy}_{\text{aa}}^{\text{mutate}}$ and $\vy_{\text{aa}}^{\text{wild}}$ are the predicted and wild-type AA types, respectively. 

While the recovered AA distribution does not always guarantee to discover the optimal evolutionary direction for any desired property in protein engineering, it is advisable to fine-tune the designed general model to better fit the protein-specific or task-specific contexts when possible. If the mutational plans of the protein are partially discovered, \ie a certain amount of labeled experimental results for mutational assays are available, the predicted probabilities can be transformed to the mutational score of interest by additional fully-connected layers $\vy_{\text{score}} = \operatorname{MLP}(\hat{\vy}_{\text{aa}}^{\text{mutate}})$ to tailor a protein- and property-specific scoring functions. To access optimal node representation, the learnable parameters in the embedding EGC layers will be updated from pre-trained results. The training target at this stage is to minimize the gap between the predicted and true distributions of the scores. We thus adopt KL-divergence to measure the discrepancy.



\subsection{Experimental Setup}
We pre-train \textsc{LGN} on \textbf{CATH v4.3.0} \citep{ORENGO19971093} with artificial noise to predict AA type and biochemical properties (SASA and B-factor). Hidden node embeddings are learned by SE(3)-equivariant graph convolutions. The performance is validated by variant effects prediction task with DMS assays \cite{fowler2014deep}. 

\paragraph{Baseline Models}
We compare our model with a diverse of zero-shot or supervised state-of-the-art models on the fitness of mutation effects prediction that are learned with protein sequences and/or structures. \textsc{DeepSequence} \citep{riesselman2018deep} \footnote{Official implementation at \url{https://github.com/debbiemarkslab/DeepSequence}} trains VAE on protein-specific MSAs to capture higher-order interactions from the distribution of an AA sequence; \textsc{MSA Transformer} \cite{rao2021msa} is a language model with aligned protein sequences of interest; \textsc{ESM-1v} \citep{meier2021esm1v} makes zero-shot mutation predictions with masked language modeling; \textsc{ESM-IF1} \citep{hsu2022esmif1} \footnote{\textsc{MSA Transformer}, \textsc{ESM-1v} and \textsc{ESM-IF1} are implemented following \url{https://github.com/facebookresearch/esm}. \textsc{ESM-1v} has 5 variants with different setups and learned parameters, for which we run the test on all the versions and take average performance on them.} predicts protein sequence with GVP \citep{jing2020learning}, a graph representation learning method for vector and scalar features of protein graphs; 
Both \textsc{Tranception} \citep{notin2022tranception} \footnote{Official implementation at \url{https://github.com/OATML-Markslab/Tranception}} and \textsc{ProGen2} \citep{nijkamp2022progen2} \footnote{Official implementation at \url{https://github.com/salesforce/progen}} leverage autoregressive language models to retrieve AA sequence without family-specific MSAs; and \textsc{ECNet} \cite{luo2021ecnet} \footnote{Official implementation at \url{https://github.com/luoyunan/ECNet}} trains and predicts mutational effects on a specific protein by a regression model that combines deep neural networks and evolutionary coupling analysis.

\paragraph{Lightweight Equivariant Graph Neural Networks (LGN)}
\textsc{LGN} is pre-trained with protein graphs generated from \textbf{CATH} \citep{ORENGO19971093}, which prepares a diverse set of proteins with experimentally determined 3D structures from the Protein Data Bank (PDB) and, where applicable, splits them into their consecutive polypeptide chains. We employ a non-redundant subset of \textbf{CATH v4.3.0} domains for model pre-training, where none of the domain pairs in the selected protein entities have more than $40$\% sequence identity over $60$\% of the overlap (\ie over the longer sequence in the protein pair of comparison). The revised \textbf{CATH} dataset contains $31,848$ protein domains, each of which is transformed into a protein graph defined in Section~\ref{sec:proteinGraph}. The transformed protein graphs on average have $150$ nodes and $1,504$ edges. We randomly choose $500$ graphs for validation and leave the remaining for the model pre-training. Random perturbations are assigned to AA types, dihedral angles, and 3D positions during the learning phase. At the validation step, the noises are fixed to guarantee stable and comparable measurements. The main architecture constitutes a stack of $6$ \textsc{EGC} layers following $1$ fully-connected layer to make predictions on the different learning tasks. On each node, the output is a vector representation consisting of $20$ probabilities of the masked AA, and optionally $1$ predicted SASA and $1$ predicted B-factors. The loss function by Equation~\ref{eq:lossFunc} guides the backward propagation with \textsc{Adam} \citep{kingma2014adam} optimizer. The model is trained with up to $300$ epochs with the initial rate set to $0.001$ and weight decay to $0.01$. The learning rate is dampened to $0.0001$ after $150$ epochs. In order to stabilize the training procedure, the gradient clipping is set to $4$.

\paragraph{Evaluation}
We evaluate model performance on $19$ \textit{in vivo} and \textit{in vitro} DMS experiments that cover $1$-site to $28$-sites mutant scores, including $13$ single-site DMS datasets (\textbf{BG\_STRSQ} \cite{romero2015dissecting}, \textbf{BLAT\_ECOLX} \cite{jacquier2013capturing}, \textbf{HG\_FLU} \cite{doud2016accurate}, \textbf{KKA2\_KLEPN} \cite{melnikov2014comprehensive}, \textbf{MTH3\_HAEAESTABILIZED} \cite{rockah2015systematic}, \textbf{PA\_FLU} \cite{wu2015functional}, \textbf{PTEN\_HUMAN} \cite{mighell2018saturation}, \textbf{YAP1\_HUMAN} \cite{araya2012fundamental}, \textbf{MK01\_HUMAN} \cite{brenan2016phenotypic},\textbf{RL401\_YEAST} \cite{roscoe2013analyses}, \textbf{SUMO1\_HUMAN} and \textbf{CALM1\_HUMAN} \cite{weile2017framework},
and \textbf{RASH\_HUMAN} \cite{adkar2012protein}) and $6$ proteins with deep mutations (\textbf{F7YBW8\_MESOW} \cite{aakre2015evolving}, \textbf{GFP} \cite{sarkisyan2016local}, \textbf{CAPSD\_AAV2S} \cite{sinai2021generative}, \textbf{DLG4\_HUMAN} and \textbf{GRB2\_HUMAN} \cite{faure2022mapping}, and \textbf{RRM} \cite{melamed2013deep}).

Essentially, each dataset provides the raw protein sequence, as well as the mutational actions and fitness scores of individual mutants. The experimentally tested structures are accessed by \textsc{AlphaFold 2} \citep{jumper2021highly}. We focus on mutations of changing AA types and exclude $0.265\%$ ($470$ out of $200,349$) mutational actions that change the length of proteins (which would remove or add AAs). The graph construction method and feature attraction process are exactly the same as we did on the training dataset, except that we do not append artificial noises onto the test proteins, \ie we assume they are noisy by nature. When there is no mutational score available for fine-tuning, we send the raw unmutated test proteins directly to the pre-trained \textsc{LGN} model and use the log-odds-ratio from Equation~\ref{eq:logRatioScore} by the predicted probabilities of AA types for suggesting the rank of deep mutations. Otherwise, when accessing a fraction of mutational scores, the model will be fine-tuned to better fit the context of the specific protein and protein property. 

Since the mutational scores on individual proteins are tested by different labs for different properties, the models' prediction performance is then evaluated on Spearman's correlation between the computational and experimental scores on all the mutation combinations, where a close-to-1 correlation indicates a better prediction performance. In addition, we supplement the true positive rate (TRP) on the top $20\%$ mutational samples for the $13$ proteins to endorse the effectiveness of single-site mutations. As the supervised learning methods significantly outperform pre-trained models in a majority of cases, we compare the TRP on the two fine-tuned models, \ie \textsc{ECNet} and \textsc{LGN+}.

\section{Conclusion}
\label{sec:conclusion}
Designing directed evolution on proteins, especially with deep mutations for functional fitness, is of enormous engineering and pharmaceutical importance. However, existing experimental methods are economically expensive, and \textit{in silico} methods demand considerable computational resources. This paper presents a lightweight supervised learning method for mutational effect prediction on arbitrary numbers of amino acids by altering the problem to denoise a protein graph. Due to rare mutational records, we pre-train our model to recover amino acid types and other important molecular properties (\ie B-factor and SASA) from randomly corrupted protein observations. It enables the model to comprehend the generic protein language and hence requires fewer efforts in learning mutations' effects for particular proteins towards desired properties at a later stage. We employ translation and rotation equivariant neural message passing layers to extract geometric-aware representation for the microenvironment of central AAs and thus grasp rich information for efficiently learning protein functions. Instead of making autoregressive interpretations along the chain, our model predicts the joint distribution of all the amino acid types all at once, enabling epistatic effects that can potentially discover better mutants than natural selection. 
Our method outperforms existing SOTA self-supervised and fine-tuned models on $19$ public proteins in predicting the deep mutational scanning assays, while consuming significantly fewer computational resources.

\backmatter

\newpage

\begin{appendices}

\section{Summary on Test Proteins}
Table~\ref{tab:testProtSummary} summarizes the characterization of the mutational scanning dataset with higher-order mutations, including the protein length and the number of scores they recorded in different orders of mutations. Note that there are around $0.265\%$ ($470$ out of $200,349$) mutational actions changing the length of proteins in the original dataset, which are removed in this work to focus on mutational actions that alter AA types.

\begin{table}[!htp]
\caption{Summary of the Higher-Order Mutation Test Dataset.}
\begin{center}
\label{tab:testProtSummary}
\resizebox{\linewidth}{!}{%
    \begin{tabular}{crrrrrr}
    \toprule
    & \textbf{CAPSD\_AAV2S} & \textbf{GFP} & \textbf{F7YBW8\_MESOW}& \textbf{DLG4\_HUMAN} & \textbf{GRB2\_HUMAN} & \textbf{RRM}  \\
    \midrule
    \# node & $734$ & $235$ & $92$& $723$ & $216$ & $77$ \\
    \midrule
    1 & $1,064$ & $1,084$ & $37$ & $1,280$ & $1,034$ & $1,064$ \\
    2 & $21,666$ & $12,777$ & $499$ & $5,696$ & $62,332$ & $36,522$ \\
    3 & $13,812$ & $12,336$ & $2,798$ \\
    4 & $13,292$ & $9,387$ & $5,858$  \\
    5 & $12,596$ & $6,825$ &  &  \\
    6 & $10,792$ & $4,298$ &  & \\
    7 & $1,716$ & $2,526$ &  & \\
    8 & $1,478$ & $1,364$ &  &  \\
    9 & $1,302$ & $627$ &  &   \\
    10 & $1,166$ & $299$ &  &  \\
    11 & $890$ & $118$ &  &  \\
    12 & $814$ & $43$  &  &  \\
    13 & $736$ & $23$  &  &  \\
    14 & $656$ & $5$   &  &  \\
    15 & $572$ & $2$   &  &  \\
    16 & $472$  &  \\
    17 & $406$  &  \\
    18 & $318$  &  \\
    19 & $238$ &  \\
    20 & $186$ &  \\
    21 & $148$ &  \\
    22 & $112$ &  \\
    23 & $86$ &  \\
    24 & $58$ &  \\
    25 & $34$ &  \\
    26 & $24$ &  \\
    27 & $16$ &  \\
    28 & $6$ &  \\ \midrule
    \textbf{sum} & $84,656$ & $51,714$ & $9,192$ & $6,976$ & $63,366$ & $37,586$ \\
    \bottomrule \\
    \end{tabular}}
\end{center}
\end{table}

\end{appendices}

\bibliography{ref}

\end{document}